\documentclass[twocolumn,superscriptaddress,showpacs,prl]{revtex4}

\usepackage[english]{babel}
\usepackage{epsfig}
\usepackage{amsmath}
\usepackage{comment}
\usepackage{graphicx}

\newcommand{\bs}{\boldsymbol}


\begin{document}

\title{Dynamical Origins for Non-Gaussian Vorticity Distributions in Turbulent Flows}

\author{Michael Wilczek}
\email{mwilczek@uni-muenster.de}
\author{Rudolf Friedrich}
\affiliation{Institute for Theoretical Physics, University of M\"unster, Wilhelm-Klemm-Str. 9, 48149 M\"unster, Germany}

\begin{abstract}
We present results on the connection between the vorticity equation and the shape of the single-point vorticity PDF. The statistical framework for these observations is cast in form of conditional averages. The numerical evaluation of these conditional averages provides insights into the intimate relation of dynamical effects like vortex stretching and vorticity diffusion and non-Gaussian vorticity statistics.
\end{abstract}

\maketitle

One of the main goals of theoretical turbulence research is to develop a statistical non-equilibrium theory of turbulence from first principles. While this goal has not been achieved so far, many phenomenological theories exist, explaining a variety of phenomena of turbulent flows \cite{frisch95book,friedrich97prl}. For example, the classical K41 \cite{kolmogorov91prs} theory manages to predict the observed energy spectrum, and later theories like K62 or the multifractal model \cite{frisch95book} even give arguments for the intermittent scaling of the structure functions for the velocity increments. However, all these classical theories do not directly link to the equation of motion of a fluid, i.e. the Navier-Stokes (or equivalently the vorticity) equation.\par 
There are, however, works attempting to derive a direct statistical theory of turbulence from first principles. For example, Lundgren \cite{lundgren67pof} derived an evolution equation for the turbulent velocity with the help of PDF methods. Similar results were obtained by Novikov \cite{novikov93jfr}, however focusing on the vorticity. Extensions to turbulent combustion and many other applications were proposed by Pope \cite{pope00book}. Mathematically all these approaches have to face the famous closure problem of turbulence, which is inherent to the highly nonlinear character of the equations of motion. In Lundgren's work this problem results in an infinite hierarchy of evolution equations, involving an ever increasing number of velocities at different spatial points. In the works by Novikov a formal closure is achieved by introducing conditional averages, which enter the statistical equations as unknown functions.\par 
It is mainly due to experimental efforts and direct numerical simulations of turbulence that today we know that turbulent flows are governed by coherent structures. In fully developed, homogeneous and isotropic turbulence these structures appear as filamentary vortex tubes. Each of these tubes generates a swirling velocity field. The ensemble of vortices present in a turbulent fluid then interacts in a nonlinear manner, leading to the complex spatio-temporal structure of turbulent flows. In the interaction of these structures, some basic dynamical mechanisms can be identified: advection, vortex stretching and viscous diffusion of vorticity.\par 
It is a scope of the present work to establish a link between these dynamical aspects of turbulence and the non-Gaussian vorticity probability functions (PDFs). To this end, we present an evolution equation for the turbulent vorticity PDF in the spirit of Lundgren and Novikov. The appearing conditional averages can be evaluated with the help of highly resolved direct numerical simulations of the vorticity equation. The results help to reveal the connections between turbulent dynamics, coherent structures and non-Gaussian statistics.\par

The temporal evolution of the vorticity $\bs \omega(\bs x,t)$ is governed by the vorticity equation,
\begin{equation}\label{eq:vorticity}
  \frac{\partial \bs \omega}{\partial t}+\bs u\cdot\nabla\bs\omega=\bs S\cdot \bs\omega+\nu\Delta\bs\omega+\bs F ,
\end{equation}
where $\bs u(\bs x,t)$ denotes the velocity field and $S_{ij}=\frac{1}{2}\left( \frac{\partial u_i}{\partial x_j}(\bs x,t)+\frac{\partial u_j}{\partial x_i}(\bs x,t) \right)$ denotes the rate-of-strain tensor. As we want to focus on incompressible fluids, the velocity field can be obtained from the vorticity field via Biot-Savart's law. $\nu$ denotes the kinematic viscosity and $\bs F(\bs x,t)$ is an external forcing in order to achieve a statistically stationary state. Starting from the fine-grained PDF $\hat f(\bs \Omega;x,t)=\delta(\bs \omega(\bs x,t)-\bs \Omega)$ standard PDF methods yield a kinetic equation for the turbulent vorticity PDF $f(\bs \Omega; \bs x, t)=\langle \hat f(\bs \Omega; \bs x, t) \rangle$ \cite{lundgren67pof,novikov93jfr,pope00book},
\begin{equation}\label{eq:kinetic1}
  \frac{\partial}{\partial t}f+\nabla_{\bs x} \cdot \lbrace \langle \bs u | \bs \Omega \rangle f \rbrace=-\nabla_{\bs \Omega}\cdot\lbrace \langle \bs S\cdot \bs \omega+ \nu \Delta_{\bs x} \bs \omega+\bs F | \bs \Omega \rangle f \rbrace .
\end{equation}
The right hand side of this kinetic equation for the vorticity PDF reveals the different dynamical influences: the average vortex stretching term, vorticity diffusion and the forcing conditioned on the sample-space vorticity. Taking into account homogeneity, the advective term vanishes, as both the conditional average as well as the PDF do not depend on the spatial coordinate. It is argued in \cite{novikov93jfr} that the conditional average of the forcing decays with increasing Reynolds number like $\langle \bs F | \bs \Omega\rangle \sim Re^{-3/2}$. Given a sufficiently high Reynolds number, the large-scale forcing should not affect the smallest scales of the flow. As the coherent structures live on these scales, a negligible influence of the forcing is physically sound. With these simplifications the kinetic equation reads
\begin{equation}\label{eq:kinetic2}
  \frac{\partial}{\partial t}f=-\nabla_{\bs \Omega}\cdot\lbrace \langle \bs S\cdot \bs \omega+ \nu \Delta_{\bs x} \bs \omega | \bs \Omega \rangle f \rbrace .
\end{equation}
In order to achieve a statistically stationary state, the right hand side of this equation has to vanish. This is only possible by a \textit{statistical} cancellation of the appearing terms. In the following we will, in addition to homogeneity, consider isotropic turbulence. This imposes further constraints on the statistical quantities. It follows that $\langle \nu \Delta \bs \omega | \bs \Omega \rangle \sim \bs \Omega$. With $\langle \bs S\cdot \bs \omega | \bs \Omega \rangle=\langle \bs S | \bs \Omega \rangle \cdot \bs \Omega$, the conditioned rate-of-strain tensor can accordingly be written down as $\langle S_{ij} | \bs \Omega \rangle=g(\Omega)\delta_{ij}-3h(\Omega)\frac{\Omega_i\Omega_j}{\Omega^2}$, with two scalar functions $h$ and $g$ only depending on the absolute value of the sample-space vorticity. The trace of this conditionally averaged tensor has to vanish, which yields $g=h$. With this it is easy to show that $\bs \Omega$ is an eigenvector of $\langle S_{ij} | \bs \Omega \rangle$, $\langle S_{ij} | \bs \Omega \rangle\Omega_j=-2g(\Omega)\Omega_i$ with the eigenvalue $\lambda_1=-2g(\Omega)$. The remaining eigenvalues can directly be determined due to the trace condition and isotropy, $\lambda_{2,3}=g(\Omega)$. Thus a cancellation of the two terms on the right hand side of (\ref{eq:kinetic2}) is possible in a statistical sense. While these observations show, which dynamical effects statistically have to cancel, these equations do not suffice to directly calculate the vorticity PDF. This can be achieved by taking into account the homogeneity of the flow. Calculating the Laplacian of the  vorticity PDF yields
\begin{equation}\label{eq:homogeneity}
\frac{\partial^2}{\partial_{x_i^2}} f=0=-\frac{\partial}{\partial \Omega_j} \langle \frac{\partial^2 \omega_j}{\partial x_i^2} | \bs \Omega  \rangle f+\frac{\partial^2}{\partial \Omega_j \partial \Omega_k}\langle \frac{\partial \omega_j}{\partial x_i} \frac{\partial \omega_k}{\partial x_i} | \bs \Omega \rangle f. 
\end{equation}
In case of a single component, this equation reduces to $0=-\frac{\partial}{\partial \Omega_x}\langle \Delta \omega_x | \Omega_x  \rangle  f- \frac{\partial^2}{\partial \Omega_x^2}\langle (\nabla\omega_x)^2 | \Omega_x \rangle f$, which suffices to determine the functional form of the vorticity PDF for homogeneous (and not necessarily stationary) flows \cite{ching96pre}.\par
An appealing kinetic equation arises, when combining the kinetic equation with the homogeneity relation (\ref{eq:homogeneity}). For homogeneous turbulent flows, the temporal evolution of the vorticity PDF can then be described by
\begin{equation}\label{eq:kinetic3}
  \frac{\partial}{\partial t}f=-\frac{\partial}{\partial \Omega_i} \langle S_{ij}\omega_j| \bs \Omega \rangle f - \frac{\partial^2 }{\partial \Omega_i \partial \Omega_j} \langle \nu \left( \frac{\partial \omega_i}{\partial x_k}\frac{\partial \omega_j}{\partial x_k} \right) | \bs \Omega \rangle f.
\end{equation}
For a single component, equation (\ref{eq:kinetic3}) reads
\begin{equation}\label{eq:kinetic3_comp}
  \frac{\partial}{\partial t}f=-\frac{\partial }{\partial \Omega_x} \langle (\bs S\cdot \bs \omega)_x | \Omega_x \rangle f - \frac{\partial^2 }{\partial \Omega_x^2} \langle \nu (\nabla \omega_x)^2 | \Omega_x \rangle f,
\end{equation}
which yields the stationary solution
\begin{equation}\label{eq:statsol}
  f(\Omega_x)=\frac{{\cal N}}{\langle \nu (\nabla \omega_x)^2 | \Omega_x \rangle}\exp\left( -\int_{-\infty}^{\Omega_x} d\Omega_x' \frac{\langle (\bs S\cdot \bs \omega)_x | \Omega_x' \rangle}{\langle \nu (\nabla \omega_x)^2 | \Omega_x' \rangle} \right)
\end{equation}
with a normalization constant ${\cal N}$. This shows, that in a stationary homogeneous flow the vorticity PDF is determined by the dynamical effect of vortex stretching and the vorticity gradient.\par
Further information on the shape of the PDF is obtained by studying the non-stationary case of equation (\ref{eq:kinetic2}) with the method of characteristics. The ordinary differential equation for the characteristics reads \mbox{$\dot {\bs \Omega}= \langle \bs S\cdot \bs \omega+ \nu \Delta_{\bs x} \bs \omega | \bs \Omega \rangle$} with the solution describing the temporal evolution of the sample-space vorticity.\par
In the following we will study the PDF $f(\Omega_x;t)$ of a single component $\Omega_x$ of the vorticity. Equation (\ref{eq:kinetic2}) then reduces to
\begin{equation}\label{eq:kinetic_comp}
  \frac{\partial}{\partial t}f=-\frac{\partial}{\partial \Omega_x}\big\lbrace \langle (\bs S\cdot \bs \omega+ \nu \Delta_{\bs x} \bs \omega)_x | \Omega_x \rangle f \big\rbrace .
\end{equation}
The statistical evolution of the sample-space vorticity $\Omega_x$ is accordingly given by $\dot \Omega_x=\langle (\bs S\cdot \bs \omega)_x + \nu \Delta_{\bs x} \omega_x | \Omega_x \rangle$,
i.e. by the sum of the conditionally averaged vortex stretching term and the conditionally averaged vorticity diffusion. These averages will be evaluated numerically below.\par
\begin{table}
\begin{center}
  \begin{tabular}{rcccccccccc}
    \hline 
    & $N$ & $R_{\lambda}$ & $u_{\mathrm{rms}}$ & $\nu$ & $L$ & $T$ & $\eta$ & $\tau_{\eta}$ & $ k_{\mathrm{max}}\eta$\\
    \hline 
    & 512 & 164 & 0.082 & 0.0001 & 2.19 & 26.73 & 0.008 & 0.63 & 1.6\\
    \hline
  \end{tabular}
\end{center}
\caption{Major simulation parameters. Number of collocation points $N$, Reynolds number based on the Taylor micro-scale $R_{\lambda}$,
  root-mean-square velocity $u_{\mathrm{rms}}$, kinematic viscosity $\nu$, integral length scale $L$, large-eddy turnover time $T$, Kolmogorov length scale $\eta$, Kolmogorov time scale $\tau_{\eta}$, $ k_{\mathrm{max}}\eta$ characterizes the spatial resolution of the smallest scales.}
\label{tab:simpara}
\end{table}
The turbulent fields under consideration in the present work are generated by a standard, dealiased Fourier-pseudo-spectral code \cite{canuto87book,hou2007jcp} for the vorticity equation. The integration domain is a triply periodic box of box-length $2\pi$. To obtain a statistically stationary flow we apply a large scale forcing. The time-stepping scheme is achieved by a third order Runge-Kutta scheme \cite{shu88jcp}.\par
For the present work, we conduct two different types of simulations. For an estimation of the conditional averages determining the stationary PDF a run in the statistically stationary regime is performed, table \ref{tab:simpara} sums up the major simulation parameters. In order to gain deeper insights into the formation of the PDF a second, nonstationary run with comparable simulation parameters is performed. The initial condition of this run exhibits a Gaussian vorticity distribution and the same energy spectrum as the statistically stationary simulation. The field evolves under the dynamics of the vorticity equation and during the relaxation to the statistically stationary state the non-Gaussian vorticity PDF emerges. \par
\begin{figure}
  \includegraphics[width=0.45\textwidth]{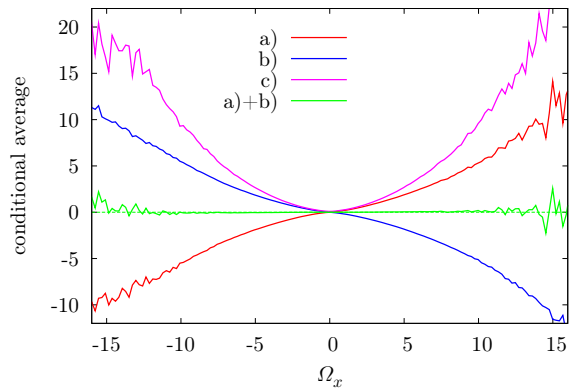}
  \caption{Conditional averages of the vortex stretching term a) $\langle (\bs S \cdot \bs \omega)_x | \Omega_x \rangle$, the vorticity diffusion b) $\langle \nu \Delta \omega_x | \Omega_x \rangle$ and the sum of both terms. The sum vanishes, as required for statistical stationarity. The squared vorticity gradient c) $\langle \nu (\nabla \omega_x)^2 | \Omega_x \rangle$ exhibits a strong dependence on $\Omega_x$.}
  \label{fig:con_cancel}
\end{figure}

\begin{figure}
  \includegraphics[width=0.47\textwidth]{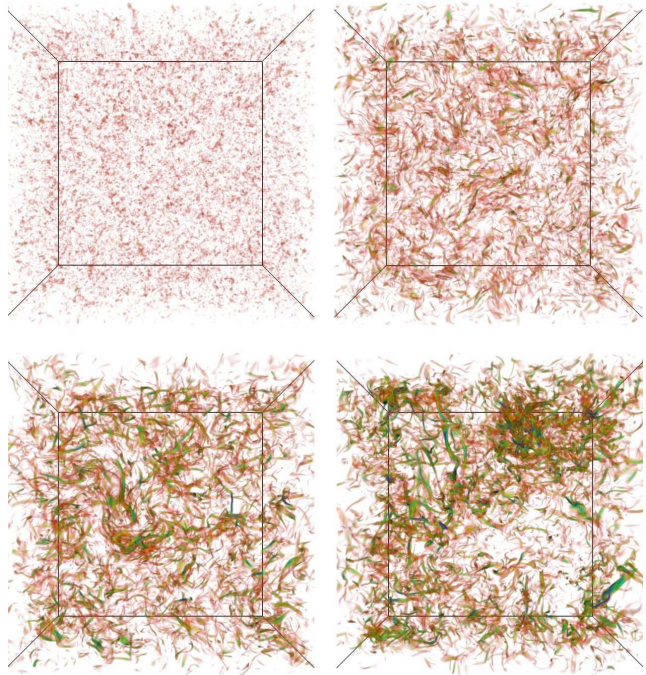}
  \caption{Volume rendering of the absolute value of vorticity above a fixed threshold for different stages of the simulation (from top left to bottom right: initial condition,  $0.11 \, T$ (200 timesteps), $0.38 \, T$ (600 timesteps), $3.53 \, T$ (6000 timesteps)).}
  \label{fig:ome_evolution}
\end{figure}

 
Regarding the stationary situation, figure \ref{fig:con_cancel} shows the numerically evaluated conditional averages of equation (\ref{eq:kinetic_comp}). The conditionally averaged vortex stretching term is positively correlated with the vorticity component, while the conditionally averaged vorticity diffusion shows strong anticorrelations. These tendencies can be physically understood; in presence of strong vorticity the vortex stretching term causes a self-amplification of vorticity, the diffusive term then tends to deplete this vorticity. 
\begin{figure}
  \includegraphics[width=0.45\textwidth]{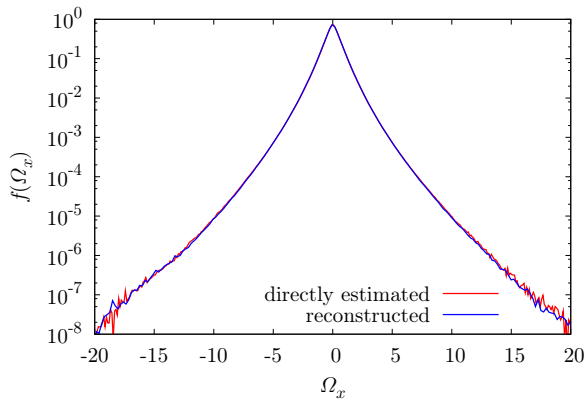}
  \caption{Logarithmic plot of the vorticity PDF estimated directly from our data and the reconstructed PDF according to eq. (\ref{eq:kinetic3_comp}). The agreement over several orders of magnitude is excellent, slight deviations are visible in the far tails of the PDF.}
  \label{fig:stat_pdf}
\end{figure}
Checking the validity of equation (\ref{eq:statsol}) with our numerical data, the comparison of the directly estimated and reconstructed vorticity PDF is depicted in \mbox{figure \ref{fig:stat_pdf}.} An excellent agreement over seven orders of magnitude is given. As equation (\ref{eq:kinetic3_comp}) yields a Gaussian in case of $\langle (\bs S\cdot \bs \omega)_x | \Omega_x \rangle\sim \Omega_x$ and $\langle \nu (\nabla \omega_x)^2 | \Omega_x \rangle \sim const.$, the highly non-Gaussian shape of the vorticity PDF can be tracked down to the strong $\Omega_x$-dependence of $\langle( \nu \nabla \omega_x)^2 | \Omega_x \rangle$. Further motivation for the functional form of the conditional averages can be given when examining the basic structures present in the flow. As the visualizations in Fig. \ref{fig:ome_evolution} suggest, the flow consists of elongated vortex tubes, which can be modeled with Burgers vortices \cite{burgers48aam}. A Burgers vortex exhibits a vorticity field according to $\bs \omega=\omega(r) \, \bs e_z=\frac{\Gamma a }{4\pi\nu}e^{-\frac{a r^2}{4\nu}} \, \bs e_z$ characterized by the strain parameter $a$ and circulation $\Gamma$. For this structure the vortex stretching term and the squared vorticity gradient are readily calculated to $\bs S \cdot \bs \omega=a\bs \omega$ and $(\nabla \omega_z)^2=\left(\frac{\partial}{\partial r} \bs \omega \right)^2=\frac{a^2r^2}{4\nu^2} \bs \omega^2$. That means, for fixed $a$, $\Gamma$ and $\nu$, the vortex stretching term is a linear function of the vorticity, whereas the squared vorticity gradient turns out to be a quadratic function of the vorticity. Thinking of turbulence as an ensemble of Burgers-like vortices, this picture already captures the main features of the conditional averages shown in Fig. \ref{fig:con_cancel}. Deviations from this simple argument are possible due to the fact that the circulation of a vortex tube in a turbulent flow is not independent of the surrounding strain field and that straight vortex tubes are not the only structures present.

\begin{figure}
  \includegraphics[width=0.45\textwidth]{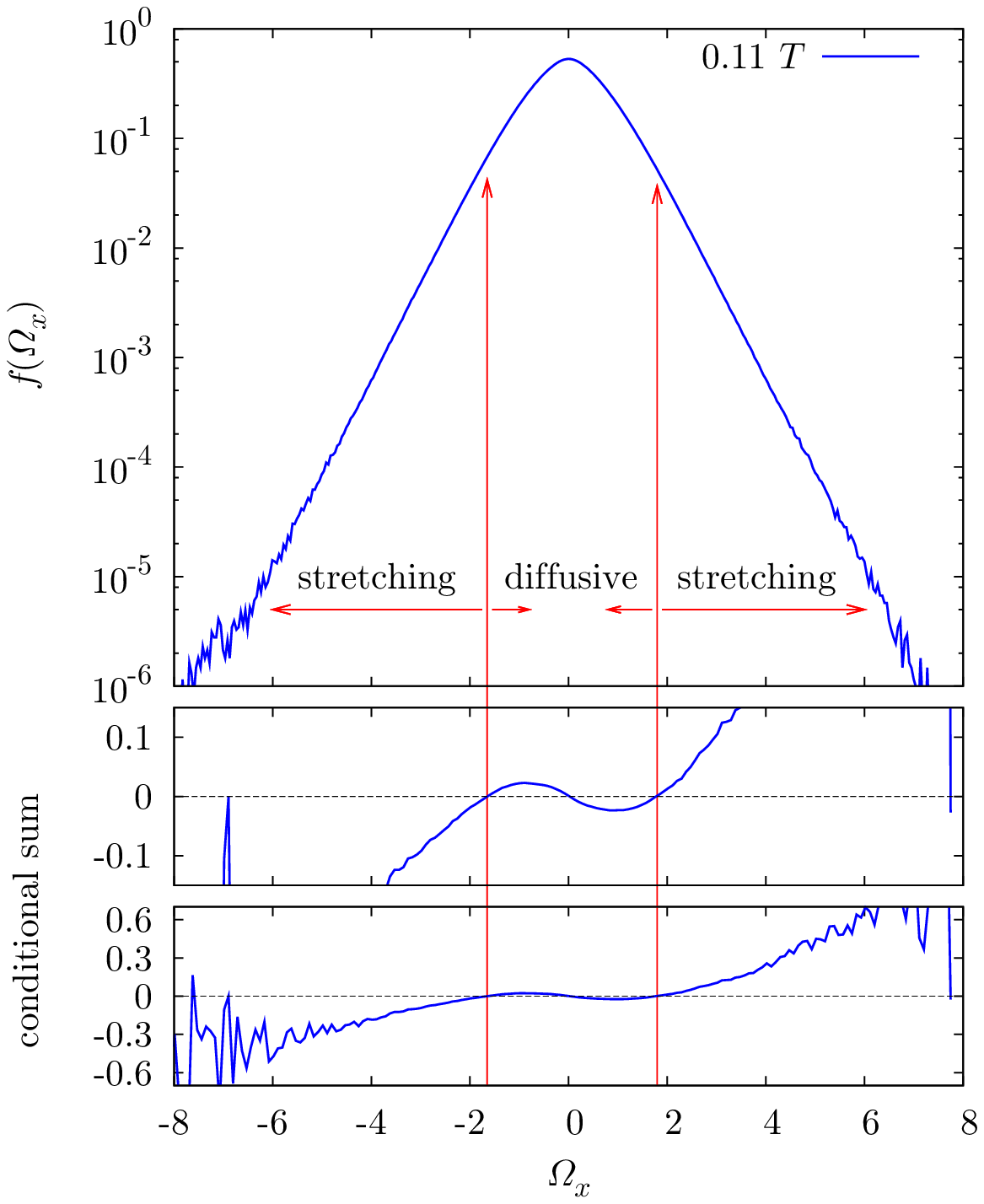}
  \caption{Illustration of the deformation of the PDF for the non-stationary simulation. Upper figure: The vorticity PDF for $0.11 T$ is already non-Gaussian, yet has not developed strong tails. Lower figures: conditional sum \mbox{$\langle (\bs S\cdot \bs \omega)_x + \nu \Delta_{\bs x} \omega_x | \Omega_x \rangle$} shown for two different y-ranges. The inner region of the vorticity PDF is quenched together due to dominant vorticity diffusion while the outer regions are stretched towards higher vorticity values due to dominant vortex stretching.}
  \label{fig:pdf_char}
\end{figure}

While helping to explain the non-Gaussian nature of the turbulent vorticity, the above considerations give no direct information on how the dynamical processes of vortex stretching and vorticity diffusion go along with the formation of the characteristic shape of the PDF. To elucidate this issue, we turn to the non-stationary simulations starting with a vorticity field exhibiting a Gaussian distribution. Under the temporal evolution of the vorticity equation (\ref{eq:vorticity}) strong spatio-temporal correlations in form of e.g. vortex tubes are generated and the initial Gaussian distribution relaxes to the non-Gaussian PDF observed for the statistically stationary regime. The temporal evolution of the vorticity field is visualized in figure \ref{fig:ome_evolution}. The initial condition with the Gaussian PDF appears unstructured, but already at $0.11 \, T$ the emergence of small vorticity worms can be observed. In the course of the simulation these structures grow stronger, after $0.38 \, T$ thin vortex tubes can be observed. Interestingly these tubes tend to cluster, as can be seen in the snapshot taken at $3.53 \, T$.\par
The temporal evolution of the sum of the conditional averages of eq. (\ref{eq:kinetic_comp}) is evaluated for the non-stationary situation. An example for $0.11 T$ is depicted in the lower part of figure \ref{fig:pdf_char}. As expected, the sum of both averages does not cancel like in the statistically stationary run. While the initial condition is purely diffusion dominated as the vortex stretching has not started generating structures, the formation of two distinct regions can be observed in the course of the simulation. Low absolute values of the vorticity are dominated by the conditional vorticity diffusion while larger absolute values are amplified due to a dominant vortex stretching. These different regions remain as the PDF is approaching stationarity, but the two terms tend to cancel more and more. After reaching the statistically stationary state the statistical balance of the conditional averages is recovered. How the unbalanced conditional sum affects the formation of the pdf may be interpreted with the method of characteristics from figure \ref{fig:pdf_char}. The inner, diffusively dominated region is quenched towards zero vorticity, while the outer, vortex stretching dominated regions are stretched towards larger absolute values of vorticity. The emerging physical picture fits well to the common understanding of turbulent flows. The stretching of the PDF towards larger absolute values of vorticity corresponds to the self-amplification of vortex filaments due to vortex stretching. The lower-valued vorticity, which corresponds to more unstructured regions of the flow is depleted by the diffusive term. Hence the formation of non-Gaussian vorticity PDF can be tracked down to the two physical mechanisms of vortex stretching and vorticity diffusion.\par

To summarize, we reported on theoretical and numerical results on the link between the vorticity equation, coherent structures and the non-Gaussian distribution of vorticity. Based on classical works by Lundgren and Novikov, an investigation of the kinetic equation of the one-point vorticity PDF reveals that the conditional averages of vortex stretching and vorticity diffusion determine the temporal evolution and shape of the vorticity PDF. A closed expression for the stationary vorticity PDF was found in terms of the conditional averages of vortex stretching and squared vorticity gradients. Numerical simulations confirm this relation with a high degree of precision. Further investigations of a non-stationary flow reveal that during the transition to the stationary state, two distinct regions of the vorticity PDF can be found: the inner region of this PDF is quenched due to dominant vorticity diffusion while the development of the fat tails can be associated with the stretching of strong vortices.\par
Hence this work highlights a direct connection between basic dynamical features of turbulence and their statistical consequences. These results encourage to study turbulent flows in terms of coherent structures as a main source for non-Gaussian statistics.\par

We acknowledge insightful discussions with F. Jenko, O. Kamps, M. Vo\ss{}kuhle and A. Daitche. Computational resources are allocated at the LRZ Munich (project h0963) and on the BOB cluster at RZ Garching. Volume rendering produced with VAPOR, \texttt{www.vapor.ucar.edu}.


\begin{thebibliography}{11}
\expandafter\ifx\csname natexlab\endcsname\relax\def\natexlab#1{#1}\fi
\expandafter\ifx\csname bibnamefont\endcsname\relax
  \def\bibnamefont#1{#1}\fi
\expandafter\ifx\csname bibfnamefont\endcsname\relax
  \def\bibfnamefont#1{#1}\fi
\expandafter\ifx\csname citenamefont\endcsname\relax
  \def\citenamefont#1{#1}\fi
\expandafter\ifx\csname url\endcsname\relax
  \def\url#1{\texttt{#1}}\fi
\expandafter\ifx\csname urlprefix\endcsname\relax\def\urlprefix{URL }\fi
\providecommand{\bibinfo}[2]{#2}
\providecommand{\eprint}[2][]{\url{#2}}

\bibitem[{\citenamefont{Frisch}(1995)}]{frisch95book}
\bibinfo{author}{\bibfnamefont{U.}~\bibnamefont{Frisch}},
  \emph{\bibinfo{title}{{T}urbulence - {T}he {L}egacy of {A}.{N}.
  {K}olmogorov}} (\bibinfo{publisher}{Cambridge University Press},
  \bibinfo{address}{Cambridge, England}, \bibinfo{year}{1995}).

\bibitem[{\citenamefont{Friedrich and Peinke}(1997)}]{friedrich97prl}
\bibinfo{author}{\bibfnamefont{R.}~\bibnamefont{Friedrich}} \bibnamefont{and}
  \bibinfo{author}{\bibfnamefont{J.}~\bibnamefont{Peinke}},
  \bibinfo{journal}{Physical Review Letters} \textbf{\bibinfo{volume}{78}},
  \bibinfo{pages}{863} (\bibinfo{year}{1997}).

\bibitem[{\citenamefont{{Kolmogorov}}(1991)}]{kolmogorov91prs}
\bibinfo{author}{\bibfnamefont{A.~N.} \bibnamefont{{Kolmogorov}}},
  \bibinfo{journal}{Proceedings of the Royal Society: Mathematical and Physical
  Sciences (1990--1995)} \textbf{\bibinfo{volume}{434}}, \bibinfo{pages}{9}
  (\bibinfo{year}{1991}).

\bibitem[{\citenamefont{Lundgren}(1967)}]{lundgren67pof}
\bibinfo{author}{\bibfnamefont{T.~S.} \bibnamefont{Lundgren}},
  \bibinfo{journal}{Physics of Fluids} \textbf{\bibinfo{volume}{10}},
  \bibinfo{pages}{969} (\bibinfo{year}{1967}).

\bibitem[{\citenamefont{Novikov}(1993)}]{novikov93jfr}
\bibinfo{author}{\bibfnamefont{E.~A.} \bibnamefont{Novikov}},
  \bibinfo{journal}{Fluid Dynamics Research} \textbf{\bibinfo{volume}{12}},
  \bibinfo{pages}{107 } (\bibinfo{year}{1993}).

\bibitem[{\citenamefont{Pope}(2000)}]{pope00book}
\bibinfo{author}{\bibfnamefont{S.}~\bibnamefont{Pope}},
  \emph{\bibinfo{title}{{T}urbulent {F}lows}} (\bibinfo{publisher}{Cambridge
  University Press}, \bibinfo{address}{Cambridge, England},
  \bibinfo{year}{2000}).

\bibitem[{\citenamefont{Ching}(1996)}]{ching96pre}
\bibinfo{author}{\bibfnamefont{E.~S.~C.} \bibnamefont{Ching}},
  \bibinfo{journal}{Phys. Rev. E} \textbf{\bibinfo{volume}{53}},
  \bibinfo{pages}{5899} (\bibinfo{year}{1996}).

\bibitem[{\citenamefont{Canuto et~al.}(1987)\citenamefont{Canuto, Hussaini,
  Quarteroni, and Zang}}]{canuto87book}
\bibinfo{author}{\bibfnamefont{C.}~\bibnamefont{Canuto}},
  \bibinfo{author}{\bibfnamefont{M.}~\bibnamefont{Hussaini}},
  \bibinfo{author}{\bibfnamefont{A.}~\bibnamefont{Quarteroni}},
  \bibnamefont{and} \bibinfo{author}{\bibfnamefont{T.}~\bibnamefont{Zang}},
  \emph{\bibinfo{title}{Spectral Methods in Fluid Dynamics}}
  (\bibinfo{publisher}{Springer-Verlag}, \bibinfo{address}{Berlin},
  \bibinfo{year}{1987}).

\bibitem[{\citenamefont{Hou and Li}(2007)}]{hou2007jcp}
\bibinfo{author}{\bibfnamefont{T.~Y.} \bibnamefont{Hou}} \bibnamefont{and}
  \bibinfo{author}{\bibfnamefont{R.}~\bibnamefont{Li}}, \bibinfo{journal}{J.
  Comp. Phys} \textbf{\bibinfo{volume}{226}}, \bibinfo{pages}{379 }
  (\bibinfo{year}{2007}).

\bibitem[{\citenamefont{Shu and Osher}(1988)}]{shu88jcp}
\bibinfo{author}{\bibfnamefont{C.}~\bibnamefont{Shu}} \bibnamefont{and}
  \bibinfo{author}{\bibfnamefont{S.}~\bibnamefont{Osher}}, \bibinfo{journal}{J.
  Comp. Phys.} \textbf{\bibinfo{volume}{77}}, \bibinfo{pages}{379 }
  (\bibinfo{year}{1988}).

\bibitem[{\citenamefont{Burgers}(1948)}]{burgers48aam}
\bibinfo{author}{\bibfnamefont{J.}~\bibnamefont{Burgers}},
  \emph{\bibinfo{title}{Advances in applied mechanics}},
  \bibinfo{howpublished}{Academic, New York} (\bibinfo{year}{1948}).

\end{thebibliography}

\end{document}